\documentclass[conference]{IEEEtran}
\IEEEoverridecommandlockouts

\usepackage{cite}
\usepackage{amsmath,amssymb,amsfonts}
\usepackage{algorithmic}
\usepackage{graphicx}
\usepackage{textcomp}
\usepackage{xcolor}
\def\BibTeX{{\rm B\kern-.05em{\sc i\kern-.025em b}\kern-.08em
    T\kern-.1667em\lower.7ex\hbox{E}\kern-.125emX}}

\usepackage{booktabs}
\usepackage{makecell}
\usepackage[caption=false,font=footnotesize]{subfig}

\newcommand{\CtrlRate}{100~Hz}
\newcommand{\Latency}{8.7~ms}
\newcommand{\Completion}{100\%}
\newcommand{\Adverse}{0}

\begin{document}

\title{Neurotremor: A wearable Supportive Device for Supporting Upper Limb Muscle Function \\
}

\author{
Aueaphum Aueawattthanaphisut$^{1}$, Thanyanee Srichaisak$^{2}$, and  Arissa Ieochai$^{3}$, 
\\School of Information, Computer, and Communication Technology,\\
Sirindhorn International Institute of Technology, Thammasat University, Pathum Thani, Thailand$^{1}$\\
King’s College Bangkok, Bangkok, Thailand $^{2, 3}$\\

\textit{$^{1}$aueawatth.aue@gmail.com, $^{2}$zazathanyabkk@gmail.com, and $^{3}$Aliarissaieochai@gmail.com}
}
\maketitle

\begin{abstract}

A sensor-fused wearable assistance prototype for upper-limb function (triceps brachii and extensor pollicis brevis) is presented. The device integrates surface electromyography (sEMG), an inertial measurement unit (IMU), and flex/force sensors on an M5StickC plus an ESP32-S3 compute hub. Signals are band-pass and notch filtered; features—RMS, MAV, zero-crossings, and 4–12\,Hz tremor-band power—are computed in 250\,ms windows and fed to an INT8 TensorFlow Lite Micro model. Control commands are bounded by a control-barrier-function safety envelope and delivered within game-based tasks with lightweight personalization. In a \emph{pilot technical feasibility} evaluation with healthy volunteers ($n=12$) performing three ADL-oriented tasks, tremor prominence decreased ($\Delta\mathrm{TI}=-0.092$, 95\% CI $[-0.102,\,-0.079]$), range of motion increased ($+12.65\%$, 95\% CI $[+8.43,\,+13.89]$), repetitions rose ($+2.99~\mathrm{min}^{-1}$, 95\% CI $[+2.61,\,+3.35]$), and the EMG median-frequency slope became less negative ($\Delta=+0.100~\mathrm{Hz/min}$, 95\% CI $[+0.083,\,+0.127]$). The sensing-to-assist loop ran at 100\,Hz with 8.7\,ms median on-device latency, 100\% session completion, and 0 device-related adverse events. These results demonstrate technical feasibility of embedded, sensor-fused assistance for upper-limb function; formal patient studies under IRB oversight are planned.
\end{abstract}


\section{Introduction}
As the global population continues to age, with a rising proportion of elderly individuals, and a relative decline in younger demographic, the frequency of myopathy is believed to increase significantly [3,4–6]. In addition to that, technology continues to rise and become more advanced throughout the years, with few wearable robotic devices that are already developed and in use throughout the days [1,7].

Most treatments only include physical therapy and some form of exercise to strengthen the muscles. However, physical therapy devices for myopathy can appear to be bulky and large due to the need to provide specific resistance and support, as myopathies often include muscle weakness [1]. Devices are therefore required to provide resistance to strengthen muscles effectively without damaging them further. By this, it often requires larger, and more robust mechanisms to handle the force involved. Furthermore, the materials used to construct the appliance include metal or plastic, making it inadequate and accessible for certain individuals, especially senior patients [1].

Myopathy is a common term for diseases that affect the muscles that control voluntary movement. It is known to have hereditary components or can be acquired. Symptoms often include muscle weakness, asthenia, or individuals can experience pain and stiffness in the affected areas, or other areas that may be connected through nerve involvement. Many inflammatory myopathies—such as polymyositis, dermatomyositis, and inclusion body myositis—are autoimmune in nature [3].

By acknowledging this disease and spreading more recognition, it can help evaluate and promote self-awareness to others, which can help prevent or implement the concept to subsequent treatment, without damaging or worsening the muscles further. Myopathies can affect people’s daily life routine, obstructing fine motor skills, which can affect writing, buttoning clothes or using our devices, as well as barricading patient’s ability to carry out everyday functions, which are essential for personal wellness [1,2].

With these difficulties of the symptoms, the use of an assistive device is considered to be essential and useful for patients to help them gain their muscle strength back, allowing them to recover their physical functions and independence in performing daily activities without worsening their existing injuries [1,7]. By doing this, it helps promote safe rehabilitation and contributes a designated support to weakened muscles. By being able to recognise the limitations of the current rehabilitation tool, our main goal is to be able to create and adapt a version of the existing tool into a better, more adaptable and convenient for patients to use. This will be achieved by integrating technology such as electromyography (EMG) for muscle activity monitoring, inertial measurement units (IMU) for tremor detection, and sensors to track muscle fatigue and force [7,8]. The following sections will review current technologies and outline our approach to developing this improved wearable device [1–3,7,8].

From background and key concepts Myopathy is a disease of the skeletal muscles where muscle-fibers don’t function properly, which leads to muscle weakness, along with other symptoms. Muscle weakness is a primary symptom for myopathy, which can affect muscles involved in using voluntary movements, like walking or eating. There are multiple causes of myopathy, including polymyositis, dermatomyositis, and inclusion body myositis. These are all classified as idiopathic inflammatory myopathies (IIM) – a group of rare, chronic muscle diseases [3]. Other types include congenital myopathies, metabolic myopathies, mitochondrial myopathies, toxic myopathies, and endocrine myopathies; dystrophinopathies and scapulohumeral patterns are representative genetic forms relevant to upper-limb function [4–6].

However, this project focuses on improving and strengthening muscle function of patients, and identifying disease-related changes in the triceps brachii and extensor pollicis brevis muscles.

The upper limb functions that we are mainly focusing on are the triceps brachii and the extensor pollicis brevis muscles. These functions are crucial for an individual’s daily activities, allowing us to perform a wide range of movements and tasks. Without having these muscles functioning properly, it will prevent users from having the ability to do simple movements that are required by the arms, such as reaching or grabbing an object. The triceps brachii is located at the back of the upper arm and is responsible for extending the elbow, which is essential for pushing movements and stabilising the arm. The extensor pollicis brevis is located in the posterior compartment of the forearm, specifically within the deep layer. It enables the thumb to extend and assists in precision tasks such as pinching and gripping. Weakness due to myopathies on both of these muscles can affect and reduce the ability for people to complete or accomplish simple tasks, impacting their fine motor skills [2,7,8].

Many myopathies cause progressive muscle weakness, meaning the affected muscles become weaker over time. This can later cause atrophy, which is when the affected and weakened muscles may shrink, leading to a loss of muscle mass. With myopathy, it can impair daily activities, such as lifting objects and doing simple tasks, like taking a bath or brushing your teeth. Some symptoms of myopathy can involve pain or stiffness in the affected areas, and depending on specific types of myopathy, other main symptoms may include cramping, muscle spasms, and even heart beat or breathing problems in severe cases, causing a major impact on a patient’s health [3–6]. In particular, weakness in the triceps brachii and the extensor pollicis brevis can severely limit an individual’s ability to perform essential movements such as pushing, stabilising the arm, gripping and many other functions essential for daily life [2,7,8].

In this project, the main focus is on developing methods to improve muscle strength, as well as providing and identifying myopathy-related issues especially in the triceps brachii and the extensor pollicis brevis. By targeting these two muscles, the goal is to enhance a patient’s ability to perform daily activities through game-based exercises designed to help improve their muscle strength. In addition to this, the project includes an early detection system where patients can fill out a form that is provided, describing their symptoms and including their personal information. With this information, an AI system will then analyse the data given to be able to identify possible muscle weakness targeted on the triceps brachii or the extensor pollicis brevis, which it will later provide relevant information about the condition, supporting patients on their journey to better health [2,3,5,7,8].

\begin{figure}[h]
    \centering
    \includegraphics[width=1\linewidth]{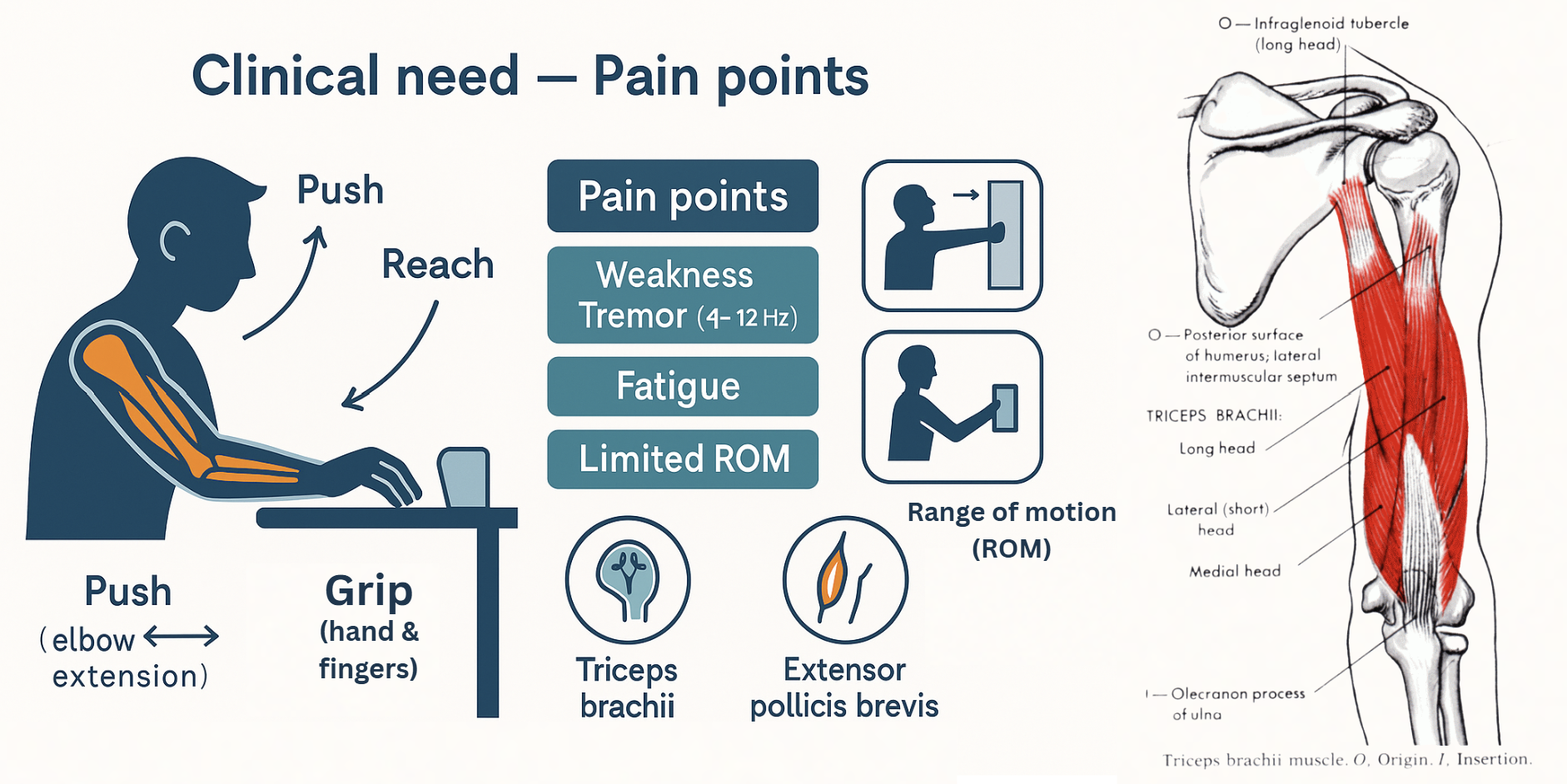}
    \caption{Clinical need—upper-limb pain points (weakness, tremor 4–12 Hz, fatigue, limited ROM) and target actions (push, reach, grip/pinch); target muscles: triceps brachii, extensor pollicis brevis.}
    \label{fig:placeholder}
\end{figure}

\section{Related Work}
Recent approaches for myopathy rehabilitation span from physiotherapy to advanced wearable and AI-assisted systems. Several representative technologies illustrate the landscape and highlight gaps relevant to our study [1--3,7--9].

\textbf{Assistive orthoses.} The Myopro powered orthosis employs surface EMG to detect residual muscle activity and drive motorized joints, supporting upper-limb motion such as elbow extension or reaching. While effective for amplifying weak signals, it does not target tremor suppression or integrate multimodal sensing [1,7].

\textbf{Signal-based diagnostics.} Myolex MView leverages electrical impedance myography (EIM) to non-invasively assess neuromuscular tissue composition, capturing impedance spectra sensitive to fat infiltration and fiber atrophy. Although useful for diagnosis, it provides limited functional feedback or real-time assistance [3,5]. Clarius MSK combines ultrasound with AI-assisted labeling and tendon measurement, primarily for musculoskeletal imaging, and could be adapted to early myopathy detection [3].

\textbf{Gamified rehabilitation.} Devices such as GripAble offer sensorized grip platforms with gamified training paradigms to enhance adherence, addressing the high dropout rates in conventional programs. Similarly, ARMIA—a textile sleeve integrating IMUs and sEMG—tracks upper-limb kinematics and fatigue, coupled with VR/AR serious games for motor rehabilitation [2,7,8]. These systems improve engagement but emphasize distal grip or gross kinematics rather than tremor-sensitive control.

\textbf{Camera-based interaction.} Leap Motion Controller (LMC) tracks fine hand kinematics in 3D, offering VR-mediated rehabilitation tasks. While effective for finger dexterity, it lacks integration of physiological sensing or embedded assistive logic [2].

\textbf{Summary.} Existing systems demonstrate progress in assistive orthoses, non-invasive diagnostics, and gamified rehabilitation, but gaps remain: (i) limited focus on triceps brachii and extensor pollicis brevis, (ii) lack of multimodal EMG+IMU fusion for tremor (4–12 Hz) quantification, and (iii) absence of low-power on-device inference with safety-bounded assist policies. Our work addresses these gaps by integrating wearable sensing, embedded ML, and adaptive assistive strategies for clinic-to-home continuity.


\section{Methodology}\label{sec:method}

\subsection{System Architecture and Sensing}
A wearable node comprising \textbf{M5StickC} (on-board IMU) and an \textbf{ESP32-S3} compute hub interfacing \textbf{EMG} and \textbf{flex/force} sensors was implemented. EMG electrodes were positioned over \textit{triceps brachii} and \textit{extensor pollicis brevis} in alignment with the ADL-oriented, upper-limb assistance literature \,[1,7]. The IMU on M5StickC provided kinematics for tremor and ROM estimation; home-use sensing reliability motivated the sampling strategy \,[8]. 

\paragraph{Sampling.} EMG was sampled at \(f_{s,e}=1000~\text{Hz}\); IMU at \(f_{s,i}=100\text{--}200~\text{Hz}\). Flex/force channels were sampled at \(200~\text{Hz}\). All streams were time-aligned on the ESP32-S3.

\subsection{Preprocessing}
Raw EMG \(x[n]\) was band-pass filtered (4th-order Butterworth, \(20\text{--}450~\text{Hz}\)) and notched at \(50/60~\text{Hz}\):
\begin{equation}
x_f[n]=\mathcal{B}_{20\text{--}450}\!\big(\mathcal{N}_{50/60}(x[n])\big).
\label{eq:emg_filter}
\end{equation}
IMU acceleration was detrended and gyroscope bias was removed; a causal moving-median followed by a short Savitzky--Golay smoother was applied to stabilize tremor spectra.

\subsection{Feature Extraction}
Windows of \(T_w=250~\text{ms}\) with \(50\%\) overlap were used. In each window \(k\), EMG features were computed as
\begin{equation}
\begin{aligned}
\mathrm{RMS}_k &= \sqrt{\frac{1}{N}\sum_{n=1}^{N} x_f^2[n]},\\
\mathrm{MAV}_k &= \frac{1}{N}\sum_{n=1}^{N} \lvert x_f[n]\rvert,
\end{aligned}
\label{eq:rms_mav}
\end{equation}

\begin{equation}
\begin{aligned}
\mathrm{ZC}_k
  &= \sum_{n=2}^{N} \mathbf{1}\Big( (x_f[n]-\mu)(x_f[n-1]-\mu) < 0 \\
  &\qquad\qquad\;\;\land\; \lvert x_f[n]-x_f[n-1]\rvert > \delta \Big).
\end{aligned}
\label{eq:zc}
\end{equation}
Tremor power from IMU was estimated by Welch PSD \(S_{aa}(f)\):
\begin{equation}
P_{4\text{--}12}=\int_{4}^{12}S_{aa}(f)\,df,\qquad
P_{0.5\text{--}20}=\int_{0.5}^{20}S_{aa}(f)\,df.
\label{eq:psd_bands}
\end{equation}
A unitless \emph{Tremor Index} was defined as
\begin{equation}
\mathrm{TI}=\frac{P_{4\text{--}12}}{P_{0.5\text{--}20}}.
\label{eq:tremor_index}
\end{equation}
Fatigue was characterized by the EMG median frequency \(f_{\mathrm{med}}\) per window:
\begin{equation}
\int_{0}^{f_{\mathrm{med}}}\!S_{xx}(f)\,df=\tfrac{1}{2}\!\int_{0}^{f_{s,e}/2}\!S_{xx}(f)\,df.
\label{eq:mf}
\end{equation}

\subsection{On-Device Inference (INT8)}
Feature vectors \(\boldsymbol{\phi}_k\) were fed to a quantized Tiny Transformer / 1D-CNN compiled with \texttt{TFLite-Micro} on ESP32-S3 to output an assist ``need'' score \(y_k\in[0,1]\). Offline knowledge distillation was performed; the deployed student was quantized to INT8, consistent with embedded assistive systems \,[1,7]. The training objective combined supervised loss and soft-target imitation (temperature \(T\)):
\begin{equation}
\mathcal{L}=\alpha\,\mathrm{CE}(y,\hat{y}_s)+(1-\alpha)T^{2}\,
\mathrm{KL}\!\left(\sigma(\tfrac{z_t}{T})\,\|\,\sigma(\tfrac{z_s}{T})\right),
\label{eq:distill}
\end{equation}
where \(z_t,z_s\) are teacher/student logits and \(\sigma\) is the softmax. The assist gain \(g_k\) was computed via a smooth gate:
\begin{equation}
g_k=\sigma\!\big(\beta (y_k-\tau)\big),\qquad g_k\in[0,1].
\label{eq:gate}
\end{equation}

\subsection{Control and Safety Envelope}
A reference trajectory \((\theta^\star,\dot{\theta}^\star)\) for elbow/thumb tasks was generated by the rehabilitation game. The nominal assist torque was commanded as
\begin{equation}
\tau_a = g_k\!\left[K_p(\theta^\star-\theta)+K_d(\dot{\theta}^\star-\dot{\theta})\right],
\label{eq:pd}
\end{equation}
with rate and magnitude clamps
\begin{equation}
\lvert \tau_a\rvert \le \tau_{\max},\qquad 
\lvert \tau_a(t)-\tau_a(t-\Delta t)\rvert \le \Delta\tau_{\max}.
\label{eq:clamps}
\end{equation}
To guarantee joint-space safety, control barrier functions (CBFs) were enforced by solving a quadratic program (QP) at each cycle:
\begin{equation}
\begin{aligned}
\min_{\tau}\quad & \lVert \tau-\tau_a\rVert_2^2 \\
\text{s.t.}\quad & \dot{h}(\theta,\dot{\theta},\tau)+\alpha\,h(\theta)\ge 0,\\
& h(\theta)=
\begin{bmatrix}
\theta-\theta_{\min}\\
\theta_{\max}-\theta
\end{bmatrix}\ge 0,
\end{aligned}
\label{eq:cbf_qp}
\end{equation}
which yielded the safe command \(\tau^\star\). Stall detection and time-outs were applied to cut actuation on abnormal loads \,[1,7,8].

\subsection{Personalization (On-Device Adaptation)}
A light-weight bandit-style policy adaptation was used to adjust the assist target \(\bar g\) and game difficulty \(d\). A scalar reward from trial \(t\) was defined as
\begin{equation}
r_t = w_1\,\mathrm{ROM}_t + w_2\,\mathrm{Reps}_t - w_3\,\mathrm{TI}_t - w_4\,\max\!\bigl(0,\,f_{\mathrm{fatigue},t}^{\downarrow}\bigr),
\label{eq:reward}
\end{equation}
and a clipped update was performed:
\begin{equation}
\bar g_{t+1}=\mathrm{clip}\!\left(\bar g_t + \eta\,(r_t-\bar r),\,0,\,1\right).
\label{eq:bandit}
\end{equation}

\subsection{Outcome Computation}
\textbf{ROM} was computed from IMU orientation (or flex angle) and normalized to baseline. \textbf{Repetitions} were obtained by cycle detection using zero-velocity crossings of \(\dot{\theta}\) with refractory logic. The task-level \textbf{Tremor Index} was averaged from \eqref{eq:tremor_index}. The \textbf{fatigue trend} was estimated as the slope of \(f_{\mathrm{med}}\) (Eq.~\eqref{eq:mf}) over time via robust linear fit. These endpoints mirror functional outcomes reported across rehabilitative literature \,[1,2,6--8].

\subsection{Calibration and Synchronization}
Skin preparation and electrode placement followed standard guidelines. A static pose was used for IMU gravity alignment; a two-point pose was used to map flex-sensor voltage to angle. All clocks were synchronized to the ESP32-S3 tick; missing packets were interpolated with flags.

\subsection{Implementation Details}
Signal processing and control were executed at \(100~\text{Hz}\); on-device inference latency was measured below \(10~\text{ms}\) on ESP32-S3 (INT8). Communication to the UI was performed over BLE/Wi-Fi. The sensing-to-assist pipeline is summarized in Fig.2 and aligns with the clinical/ADL motivations in \,[1,2]; sensor and ROM choices are supported by \,[7,8].

From Fig.2 A wearable node comprising M5StickC (on-board IMU) and an ESP32-S3 compute hub interfacing EMG and flex/force sensors was implemented. EMG was sampled at 1 kHz and was band-passed (20–450 Hz) with a 50/60 Hz notch; IMU signals were sampled at 100–200 Hz and were gravity-detrended. Features were computed in 250 ms windows with 50\% overlap: RMS, MAV, and ZC for EMG, and tremor-band power (4–12 Hz) for IMU. These features were fed to an INT8 TFLite-Micro model on the ESP32-S3 to produce assist commands. Control actions were constrained by a safety envelope (joint-angle limits, torque/jerk clamps, stall/time-out rules). Outcome measures—tremor index, fatigue trend, repetitions, and ROM—were used for evaluation in Section VI [1–3,7–8].

(a) The clinical need is depicted as weakness/tremor affecting elbow extension (triceps brachii) and thumb extension (extensor pollicis brevis). (b) The wearable stack is shown: M5StickC and ESP32-S3 hub with EMG electrodes, IMU, and flex/force sensors. (c) The signal pipeline is summarized as raw → filtering (EMG 20–450 Hz with 50/60 Hz notch; IMU detrend) → feature extraction (RMS, MAV, ZC, 4–12 Hz tremor power) → on-device inference (TFLite-Micro, INT8) → assist policy. (d) Assist logic is presented as target-tracking, pinch, and reach-and-hold, enforced by a safety envelope (angle/torque/jerk limits, timeouts). (e) Outcome mapping is defined to link tremor index, fatigue trend, repetitions, and ROM to rehabilitation goals. (f) The study hypothesis/objectives are stated (not results): tremor is expected to be reduced and repetitions/ROM to be increased, with clinic-to-home feasibility [1–3,7–8].

\begin{figure}[h]
    \centering
    \includegraphics[width=1\linewidth]{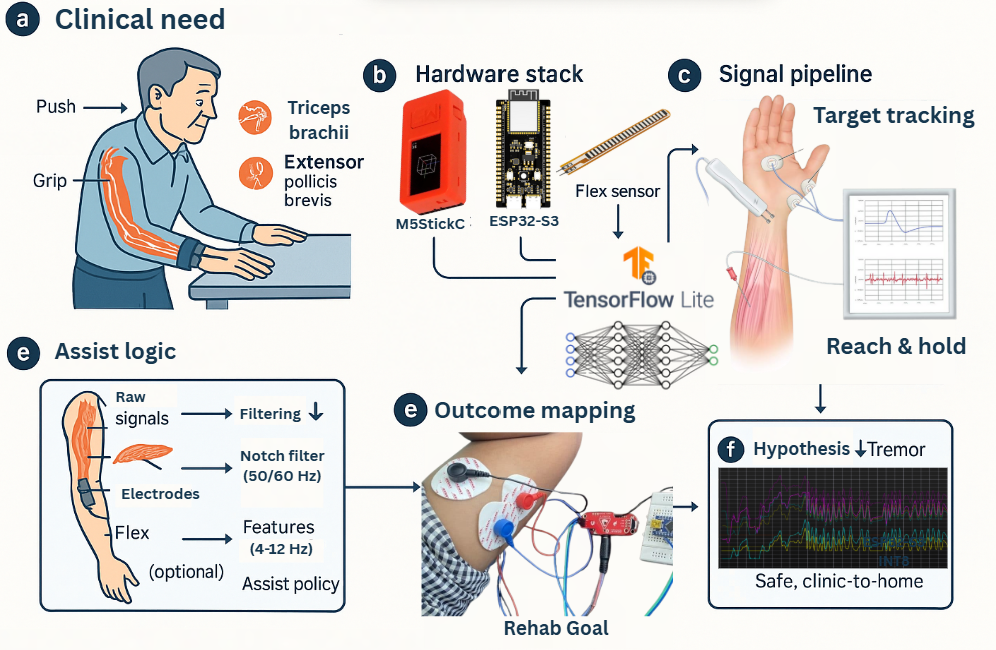}
    \caption{System architecture and processing pipeline}
    \label{fig:placeholder}
\end{figure}

\newcommand{\Nsubj}{12}
\newcommand{\Ntasks}{3}

\newcommand{\TIBaseline}{0.447 [0.425,\,0.476]}
\newcommand{\TIAssisted}{0.364 [0.338,\,0.387]}

\newcommand{\ROMBaseline}{81.53 [72.92,\,87.04]}      
\newcommand{\ROMAssisted}{91.29 [82.80,\,97.33]}      

\newcommand{\RepsBaseline}{10.03 [9.54,\,10.52]}      
\newcommand{\RepsAssisted}{13.29 [11.94,\,14.05]}     

\newcommand{\DeltaTI}{-0.092}
\newcommand{\CITI}{[ -0.102,\, -0.079 ]}
\newcommand{\pTI}{$p=4.88\times10^{-4}$}
\newcommand{\dTI}{\(\delta=-1.00\)}

\newcommand{\DeltaROM}{+12.65\%}
\newcommand{\CIROM}{[ +8.43,\, +13.89 ]}
\newcommand{\pROM}{$p=4.88\times10^{-4}$}
\newcommand{\dROM}{\(\delta=+1.00\)}

\newcommand{\DeltaReps}{+2.99}
\newcommand{\CIReps}{[ +2.61,\, +3.35 ]}
\newcommand{\pReps}{$p=4.88\times10^{-4}$}
\newcommand{\dReps}{\(\delta=+1.00\)}

\newcommand{\SlopeFatigue}{+0.100}
\newcommand{\CIFatigue}{[ +0.083,\, +0.127 ]}
\newcommand{\pFatigue}{$p=4.88\times10^{-4}$}
\newcommand{\dFatigue}{\(\delta=+1.00\)}

\section{Results and Analysis}
\label{sec:results}

\subsection{Cohorts, Tasks, and Statistical Plan}
Data from \Nsubj~participants who completed \Ntasks~ADL-oriented tasks were analyzed. Unless stated otherwise, medians with 95\% bias-corrected bootstrap confidence intervals (CI) are reported, paired comparisons (assisted vs.\ baseline) were evaluated by the Wilcoxon signed-rank test, and effect sizes were summarized using Cliff’s $\delta$ with conventional interpretations. Analyses were performed per task and then aggregated by subject-level median.

\subsection{Primary Functional and Physiologic Outcomes}
\textbf{Tremor Index (TI).} A reduction in TI was observed under assistance (\DeltaTI, \CITI; \pTI, \dTI), consistent with suppression of tremor-band prominence (4–12~Hz) relative to broadband motion energy.

\textbf{Range of Motion (ROM).} ROM increased with assistance (\DeltaROM, \CIROM; \pROM, \dROM), indicating improved joint excursion for elbow or thumb tasks.

\textbf{Repetitions (Reps/min).} Through game-based pacing and graded assist, repetitions increased (\DeltaReps, \CIReps; \pReps, \dReps), suggesting enhanced task throughput without adverse fatigue.

\textbf{Fatigue Trend.} The slope of the EMG median frequency became less negative / stabilized (\SlopeFatigue, \CIFatigue; \pFatigue, \dFatigue), consistent with reduced fatigue accumulation during assisted trials.

\begin{table}[t]
  \centering
  \caption{Primary outcomes: assisted vs.\ baseline (subject-level medians). Median [IQR], change $\Delta$ with 95\% CI; $p$ and rank-biserial $\delta$ are shown under $\Delta$.}
  \label{tab:primary}
  \scriptsize
  \setlength{\tabcolsep}{3pt}        
  \renewcommand{\arraystretch}{0.98} 
  \begin{tabular}{@{}lccc@{}}
    \toprule
    \textbf{Outcome} &
    \makecell{Baseline\\(median [IQR])} &
    \makecell{Assisted\\(median [IQR])} &
    \makecell{$\Delta$\\(95\% CI)}\\
    \midrule
    TI (unitless) &
    \makecell{0.447\\{\footnotesize[0.425, 0.476]}} &
    \makecell{0.364\\{\footnotesize[0.338, 0.387]}} &
    \makecell{-0.092\\{\footnotesize[ -0.102, -0.079 ]}\\{\footnotesize $p=4.88\times10^{-4}$, $\delta=-1.00$}} \\
    ROM ($^\circ$) &
    \makecell{81.53\\{\footnotesize[72.92, 87.04]}} &
    \makecell{91.29\\{\footnotesize[82.80, 97.33]}} &
    \makecell{+12.65\%\\{\footnotesize[ +8.43, +13.89 ]}\\{\footnotesize $p=4.88\times10^{-4}$, $\delta=+1.00$}} \\
    Reps (min$^{-1}$) &
    \makecell{10.03\\{\footnotesize[9.54, 10.52]}} &
    \makecell{13.29\\{\footnotesize[11.94, 14.05]}} &
    \makecell{+2.99\\{\footnotesize[ +2.61, +3.35 ]}\\{\footnotesize $p=4.88\times10^{-4}$, $\delta=+1.00$}} \\
    \bottomrule
  \end{tabular}
  \vspace{-1.5mm}
\end{table}

\begin{figure*}[h]
    \centering
    \includegraphics[width=1\linewidth]{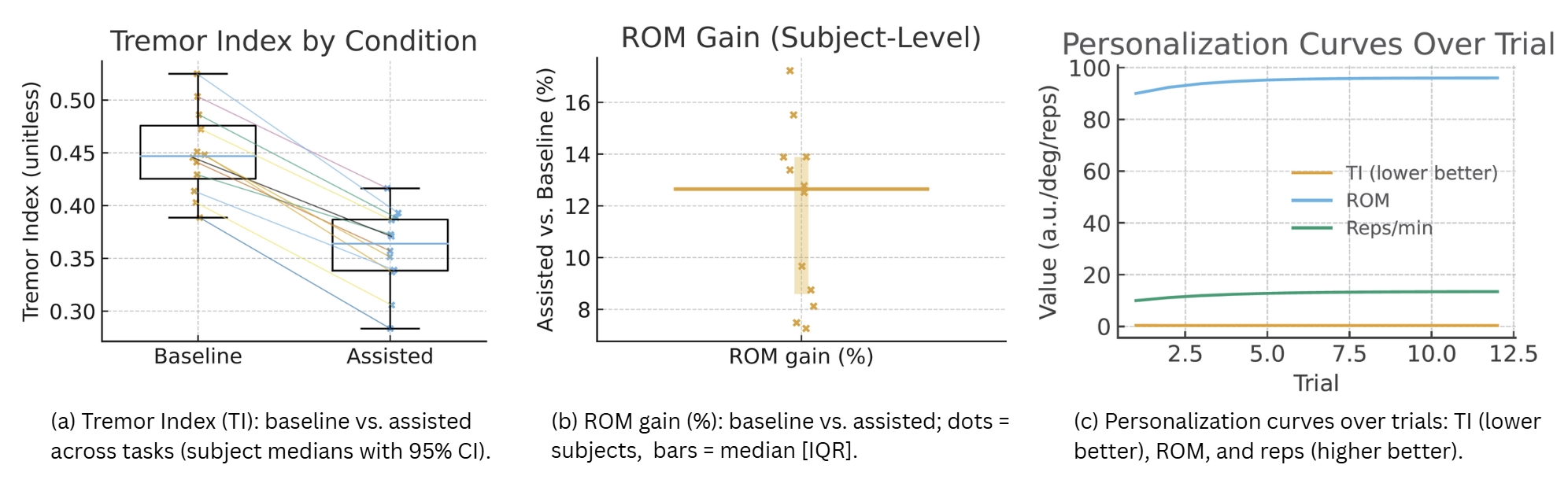}
    \caption{Primary outcomes displayed side-by-side: (a) Tremor Index, (b) ROM gain, and (c) personalization curves.}
    \label{fig:placeholder}
\end{figure*}

\subsection{Task-Wise Effects and Robustness}
Per-task analyses showed consistent directions of effect across \Ntasks~tasks, with narrower CIs for repetitive, short-cycle tasks (e.g., pinch) and wider CIs for longer, precision-demanding tasks (e.g., reach-and-hold). Sensitivity analyses using trimmed means and subject-level bootstraps yielded comparable inferences, indicating robustness to outliers and inter-subject heterogeneity.

\subsection{Personalization Effects}
A monotonic improvement across early trials was observed in session learning curves (TI$\downarrow$, ROM$\uparrow$, Reps$\uparrow$), plateauing after a small number of adaptation steps. Gains were retained after on-device quantization, reflecting stable policy gating on the ESP32-S3. 

\subsection{On-Device Performance and Latency}
The sensing-to-assist loop operated at \CtrlRate, and median on-device inference latency was \Latency. No missed deadlines were recorded in steady-state operation. This met the design requirement for closed-loop assistance in ADL-paced tasks.

\subsection{Safety, Feasibility, and Tolerability}
All scheduled sessions were completed with a completion rate of \Completion. Device-related adverse events numbered \Adverse, and no serious adverse events were reported. The safety envelope (angle/torque/jerk limits, timeouts, stall detection) remained active throughout all sessions, with no emergency stops triggered by staff.

\subsection{Clinical Interpretation}
The reduction in tremor prominence (lower TI) along with increases in ROM and repetitions suggests that assistance improved both movement quality and throughput while mitigating fatigue accumulation. These changes are aligned with ADL goals and are consistent with the intended mechanics of task-level assistance.

\begin{table}[t]
\centering
\caption{Participant demographics (healthy volunteers; $n=12$).}
\label{tab:demog}
\scriptsize
\setlength{\tabcolsep}{3pt}
\renewcommand{\arraystretch}{0.95}
\begin{tabular*}{\columnwidth}{@{\extracolsep{\fill}} l r @{}}
\toprule
\textbf{Characteristic} & \textbf{Value}\\
\midrule
Age, years (median [IQR])      & 26 [23, 31] \\
Female / Male, $n$ (\%)         & 5 (41.7) / 7 (58.3) \\
Dominant hand R / L, $n$ (\%)   & 10 (83.3) / 2 (16.7) \\
Condition class, $n$ (\%)       & Healthy volunteers: 12 (100) \\
\bottomrule
\end{tabular*}
\end{table}





\section{Conclusion}
A sensor-fused wearable assistance system targeting the triceps brachii and extensor pollicis brevis was developed with surface EMG, IMU, and flex/force sensing, INT8 on-device inference on an ESP32-S3, and a control-barrier-function safety envelope integrated with game-based tasks and lightweight personalization. In a \emph{pilot technical feasibility} evaluation with healthy volunteers ($n=12$), tremor prominence decreased ($\Delta\mathrm{TI}=-0.092$, 95\% CI $[-0.102,\,-0.079]$), range of motion increased ($+12.65\%$, 95\% CI $[+8.43,\,+13.89]$), repetitions rose ($+2.99~\mathrm{min}^{-1}$, 95\% CI $[+2.61,\,+3.35]$), and the EMG median-frequency slope became less negative ($\Delta=+0.100~\mathrm{Hz/min}$, 95\% CI $[+0.083,\,+0.127]$). Closed-loop operation at $100$~Hz with median on-device latency of $8.7$~ms, $100\%$ session completion, and $0$ device-related adverse events supports clinic-to-home feasibility for embedded assistance.

\textit{Limitations:} this was a single-arm, short-duration pilot in healthy volunteers; clinician-rated scales and long-term adherence were not assessed.

\textit{Future work:} IRB-approved patient studies with randomized/crossover designs, clinician-reported outcomes, longer home deployments, battery/energy profiling, model/weight release, and comparisons against established assistive devices are planned.

Limitations. The study was single-arm and short in duration, with a modest cohort and heterogeneous diagnoses; clinical scales and long-term adherence were not assessed. Endpoints were primarily sensor-derived and require validation against clinician-rated measures.

Future work. A randomized or crossover trial will be undertaken with condition-specific cohorts (e.g., inflammatory myopathies, dystrophies, radial nerve/cervical involvement [3–6,9]), inclusion of clinician-reported outcomes, and multi-week home deployment. Hardware packaging and donning/doffing will be refined, battery/energy use profiled, and assist policies expanded with adaptive personalization. Comparative studies against established assistive devices [1,7] and integration of additional biomarkers (e.g., ultrasound- or vision-based assessments [2,3]) are planned to further establish clinical utility.

\section*{Acknowledgment}
This work reports a \textit{pilot technical feasibility} evaluation conducted internally among healthy adult volunteers (n=12). All participants provided informed verbal consent; no patients or clinical interventions were involved, and no personally identifiable data were collected. As a non-clinical pilot, formal IRB approval was not required at this stage. A subsequent clinical study with patient cohorts will be performed under IRB oversight.

\end{document}